# Methodology for Detecting Cyber Intrusions in e-Learning Systems during COVID-19 Pandemic

Ivan Cvitić[1,*][0000-0003-3728-6711], Dragan Peraković[1][0000-0002-0476-9373], Marko Periša[1][0000-0002-1775-0735] and Anca D. Jurcut[2][0000-0002-2705-1823]

[1] University of Zagreb, Faculty of Transport and Traffic Sciences, Vukelićeva 4, 10000 Zagreb, Croatia
[2] University College Dublin, School of Computer Science and Informatics Belfield, Dublin 4, Ireland

Corresponding author's e-mail: ivan.cvitic@fpz.unizg.hr

**Abstract.** In the scenarios of specific conditions and crises such as the coronavirus pandemic, the availability of e-learning ecosystem elements is further highlighted. The growing importance for securing such an ecosystem can be seen from DDoS (Distributed Denial of Service) attacks on e-learning components of the Croatian e-learning system. The negative impact of the conducted attack is visible in numerous users who were prevented from participating in and implementing the planned teaching process. Network anomalies such as conducted DDoS attacks were identified as one of the crucial threats to the e-learning systems. In this paper, an overview of the network anomaly phenomenon was given and botnets' role in generating DDoS attacks, especially IoT device impact. The paper analyzes the impact of the COVID-19 pandemic on the e-learning systems in Croatia. Based on the conclusions, a research methodology has been proposed to develop a cyber-threat detection model that considers the specifics of the application of e-learning systems in crisis, distinguishing flash crowd events from anomalies in the communication network. The proposed methodology includes establishing a theoretical basis on DDoS and flash crowd event traffic, defining a laboratory testbed setup for data acquisition, development of DDoS detection model, and testing the applicability of the developed model on the case study. The implementation of the proposed methodology can improve the quality of the teaching process through timely DDoS detection and it gives other socio-economic contributions such as developing a specific research domain, publicly available dataset of network traffic, and raising the cyber-security of the e-learning systems.

## Declarations

**Funding:** This research is funded by the University of Zagreb through the Grants for core financing of scientific and artistic activities of the University of Zagreb in the academic year 2019/2020 under the project (555-1) "Challenges of Information and Communication Networks and Technologies, services, and user equipment in establishing the Society 5.0 environment."
**Conflicts of interest/Competing interests:** Not applicable
**Availability of data and material:** Not applicable
**Code availability:** Not applicable
**Authors' contribution:** Conceptualization: Ivan Cvitić and Dragan Peraković; Methodology: Ivan Cvitić, Dragan Peraković, Marko Periša and Anca D. Jurcut; Formal analysis: Ivan Cvitić, Dragan Peraković, Marko Periša and Anca D. Jurcut; Investigation: Ivan Cvitić; Supervision: Dragan Peraković; Visualization: Ivan Cvitić and Marko Periša; Writing – original draft: Ivan Cvitić; Writing – review & editing: Dragan Peraković, Marko Periša and Anca D. Jurcut

## 1 Introduction

Anomaly represents samples in the data that deviate from the previously defined normal behavior of the observed phenomenon. Observed from the aspect of information and communication system, anomalies in communication, i.e. network traffic, are generally generated by one or several network devices. This is often a result from illegitimate network activities in the system, with the anomalies of network traffic having the potential of negatively affecting the operation of the information and communication system or services [1]. One of the frequent causes of anomalies in the network traffic is DDoS (Distributed Denial of Service) attack. Over the last two decades numerous studies have been directed to the development of methods, models and systems that can detect DDoS traffic in real time. Nevertheless, the number of DDoS attacks and the amount of DDoS traffic is constantly increasing, which why further research in the area of the detection of security threats of this kind is necessary [2]. Despite continuous research of network traffic anomalies, cyber attacks such as DDoS attacks are still frequent and can have numerous negative effects on the predicted performance of Information and Communication (IC) systems and IC service availability. The coronavirus pandemic (SARS-CoV-2) highlighted the importance of the availability of e-learning systems and services. E-learning systems as part of an academic open approach are exposed to security threats and feature numerous vulnerabilities and are often the target of cyber attacks [5]. This problem is further highlighted by the increased need to use e-leraning as a result of the COVID-19 pandemic. Cyber threats and attacks are still present, but the current pandemic points to the need to intensify research on detecting cyber attacks in e-learning systems because it has been observed that cyber attacks in crises have a potentially far greater negative impact than under normal circumstances.

This research's motivation came from a DDoS attack conducted on AAI@EduHr, an authentication service for e-learning users in the Republic of Croatia that occurred in March 2020 while the education process migrated online during coronavirus caused lockdown [3]. The problem of DDoS attacks on e-learning systems during coronavirus pandemic was identified by Kaspersky as well, where they noticed an increase from 350% to 550% of such attacks compared to the same month in 2019 [4]. Accordingly, they emphasized the importance of researching DDoS detection and mitigation on e-learning systems during the pandemic like crises where e-learning has no alternative for conducting the education process. This research aims to propose a research methodology for developing DDoS traffic detection models at the attack target in a scenario where flash crowd events-generated traffic represents legitimate traffic, such as e-learning services during the COVID-19 pandemic. Special focus in this paper is on the e-learning system cyber security, which, with the advent of the COVID-19 pandemic, has become the primary way of conducting teaching and other administrative processes.

The rest of the paper is organized as follows: in Section 2 related research was analyzed in DDoS attack detection and flash crowd distinguishing. Section 3 gives an overview of the network traffic anomalies from the aspect of DDoS attacks as one of the most disruptive way of IC resources availability attacks. At the end of the Section the impact of COVID-19 pandemic on the availability of e-learning systems with case studies in the Republic of Croatia is analyzed. The analysis considers and discusses the problem of flash crowd appearance and DDoS in the same scenario. Flash crowd represents a legitimate event or series of legitimate events or network traffic unusually high intensity that can disrupt the availability of online service or communication resources. In Section 4 a research methodology for network traffic anomalies detection model development is proposed. Such a model would be appropriate in the crisis scenarios such as the current pandemic where the flash crowd and DDoS attack phenomena occur in the same scenario on e-learning system without an alternative for conducting the educational process.

## 2   Related work

Previous studies define several approaches to DDoS traffic detection. They can be generally divided into two basic categories; those based on the samples and those based on the anomalies [6]. Research [7], apart from the previous ones, identifies also the approach based on entropy, and research [8] mentions the possibilities of applying the hybrid approach to DDoS traffic detection. The methods based on the sample apply the comparison of the incoming traffic with the pre-defined profiles and samples of the known network anomalies [9]. The detection of DDoS attack based on the sample can be carried out in three ways; based on the known attack signature, based on the rule (if-then), and based on the condition and transition [10]. The advantage of this detection method is a high rate of detection of the already known DDoS attacks with a low number of false positive and false negative results. The drawback is the impossibility of detecting new and unknown attacks, i.e. those not included in the database, which is used for comparison with the incoming traffic samples. Because of the problem area dynamics, the detection methods must detect unknown samples of DDoS traffic [6].

On the contrary, the approach based on the detection of the network traffic anomalies uses the pre-defined models of normal traffic which are then used to compare the incoming traffic [8]. This approach to detection has been developed to overcome the drawbacks of the detection approach based on the samples [7]. If the incoming traffic significantly differs from the defined model of normal traffic, then the incoming traffic is identified as anomaly, i.e. DDoS traffic [11]. The advantage of detecting network traffic anomalies related to the detection based on samples is the possibility of discovering unknown attacks. The main drawback of detection based on anomalies is determining the threshold values between normal traffic and anomaly [8, 12]. The anomalies of network traffic are detected when the current traffic flow values or other selected parameters exceed the pre-defined threshold of the normal traffic model. A low defined threshold results in a large number of false positive results, and high defined threshold leads to a large number of false negative results [13]. Numerous scientific methods have been used for the detection of DDoS traffic [14]. The current academic literature most frequently applies the statistical methods, machine learning methods, and softcomputing methods [14–16]. In today's big data concept environments, high cyber attack detection and response performances are demonstrated by the machine learning and artificial intelligence methods. The impact of botnet (network of infected end devices) in generating DDoS traffic is analyzed with the focus on the rising trends and widespread application of IoT devices and the impact of DDoS traffic on the e-learning system in the Republic of Croatia during COVID-19 pandemic.

The pandemic impact on e-leaning was already studied in [17], where the authors concluded that the pandemic acts as a boost for e-learning development. The importance of e-learning system in education during the COVID-19 pandemic and similar crisis situations is well proven in several studies, such as [18–20]. Because of the crucial role of e-learning, the research in preventing and reacting to cyber threats directed towards such a system needs to be intensified.

## 3  Availability of e-learning system in pandemic scenario

### 3.1  Overview of network traffic anomaly phenomenon

The anomaly represents patterns in the data that do not correspond to the observed phenomenon of the previously defined normal behavior. Data are generally generated from one or more processes that may reflect the system activity or observations collected about entities [21]. Unusual behavior of the data-generating process creates anomalies in the data. Thus, an anomaly in data often contains useful information about abnormal characteristics of a system or entity that affect the data generation process. Recognizing such features that are uncommon provides useful insights into a variety of applications. Examples of anomaly detection application are visible through numerous studies in different areas of application [21, 22]. Many IC systems collect different data types about the operating system operation, network traffic, or user activity or phishing e-mails [23]. Such data may indicate unusual behavior, i.e. anomalies caused by illegitimate activities. Credit card fraud is becoming more common due to the greater ease with which sensitive data such as credit card numbers can be compromised. In many cases unauthorized use of a credit card can show different patterns, such as a sudden increase in spending from certain locations (buying sprees) or very large transactions. Such forms can be used to detect extreme values in credit card transaction data. Sensors are often used to monitor various environmental and location parameters in many real-world applications. Sudden changes in the basic patterns can be events of interest. Event detection is one of the primary motivational applications in the field of sensor networks. In a number of medical applications, data are collected using a variety of devices such as Magnetic Resonance imaging (MR), Positron Emission Tomography (PET), or time series Electrocardiograms (ECG). Unusual patterns or anomalies in such data reflect the patient's disease condition.

In the above application examples, a model of normal data is defined from the collected data, where the anomalies are identified as deviations from such a model. Certain applications such as the detection of unauthorized intrusion into the IC system or the detection of fraud anomalies correspond to a sequence of multiple data points instead of a single data point. Sequence specificity is relevant for identifying events that represent an anomaly. Such anomalies are also called collective anomalies in the literature because they can be collectively derived from a set or sequence of data points.

Detection of anomalies in the data generally gives two types of results (outcomes): the level of the anomaly and the binary indication. Most anomaly detection methods evaluate the anomaly level of each data point. The obtained levels can be used to rank the data points according to their tendency to become anomalies. Binary notations indicate whether a data point is an anomaly or not. Anomaly levels can also be transformed into binary notations using cut-off values that are selected based on, for example, the statistical distribution of levels assigned to all the data in the observed set.

Figure 1 shows an example of data anomalies in a two-dimensional dataset. The data have two normal regions labeled $N_1$ and $N_2$ since most of the observations lie in these regions. Points that are far enough from the marked regions, such as $o_1$ and $o_2$ and the set of points in the $O_3$ region represent anomalies [24].

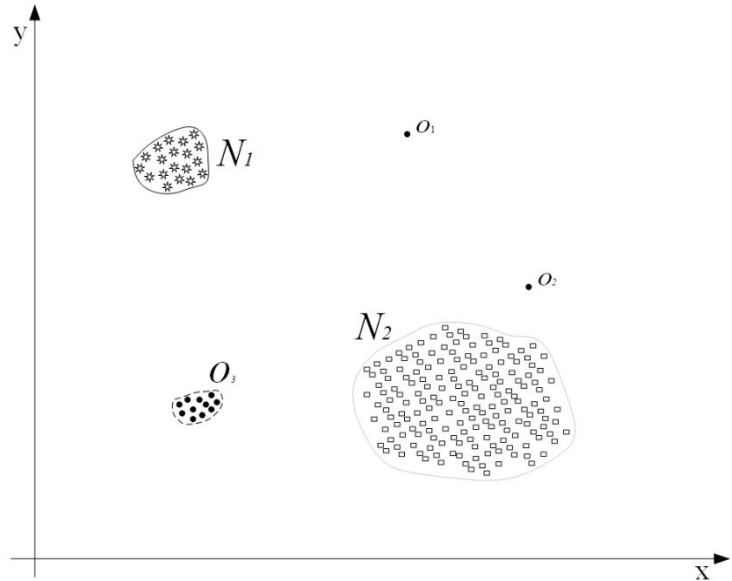

Figure 1 - Representation of an anomaly in a two-dimensional dataset [24]

Observed from the communication network aspect, anomalies can occur in the network data traffic that occurs as a result of the communication activities of terminal or network devices. According to [25], the causes of anomalies in the communication network can occur due to unexpected communication activities, a sudden increase in the demand for legitimate users' service (flash crowd) or illegitimate activities. According to [26], anomalies can be observed through three basic categories, (1) individual anomalies, (2) contextual anomalies, and (3) collective anomalies.

Individual anomalies are those in which the individual data represent an anomaly in relation to other data. These are the simplest form of anomaly and a number of anomaly detection solutions presuppose this type of anomaly. The term contextual anomaly refers to the deviation of features that represent data of a given context. The context can be temporal, spatial, or given depending on the problem area [27]. According to [28] and [29], a collective anomaly represents groups of data that behave normally at the individual level, but as a group represent an anomaly.

Detection of various anomalies caused by illegitimate activities and security incidents in the computer network is one of the most significant challenges for researchers, but also for the network administrators and security experts. One of the growing causes of network traffic anomalies is, according to numerous studies, the DDoS attack. DDoS attacks have been the subject of research since the beginning of 2000, when Yahoo, Ebay, Amazon and similar portals owned by multinational companies became targets, that even then invested large resources in the security of their IC systems. DDoS attacks already then showed that this is a type of attack that is difficult to detect and from which it is difficult to protect oneself [30, 31].

Most illegitimate activities that take place in the IC system seek to violate one of the three basic principles of security, where the violation of the principle of availability means denying access to the IC service, system, or any resource to legitimate users [32]. In doing so, a legitimate user is presented as a user who has a certain level of authority to use the IC system resources and does not exceed that authority by unauthorized activities. A Denial of Service (DoS) attack is a general class of attacks aimed at the availability of IC resources.

According to the method of distribution, DoS attacks can be divided into two categories [33]: single source attacks, Single Source Denial of Service (SDoS) and multi-source attacks or Distributed Denial of Service attack (DDoS). The source of an SDoS attack is a single computer or device on the network. A DDoS is a coordinated illegitimate activity aimed at exploiting the available capacity of devices to process or transmit network traffic in order to prevent access to IC resources to legitimate users. For the purpose of carrying out an attack, a large number of devices in the network are used (most often terminal devices) to generate illegitimate DDoS traffic to the target of the attack.

The UML state and transition diagram in Figure 2 shows the states in which the IC system or service may be located as well as the conditions that lead to the transition to individual states or retention of the existing state. Normal operation is a state that is maintained in conditions where the intensity of requests ($\lambda$) is less than the server capacity or less than the processing speed of incoming requests ($\mu$). If the request intensity is higher than the server capacity, the system enters a congestion state and remains in that state until the server queue occupancy reaches a maximum or until the intensity of incoming requests decreases. In the first case, when the maximum filling of the serving tail is reached, the system goes into a state of unavailability, while in the second case, the system returns to the state of normal operation.

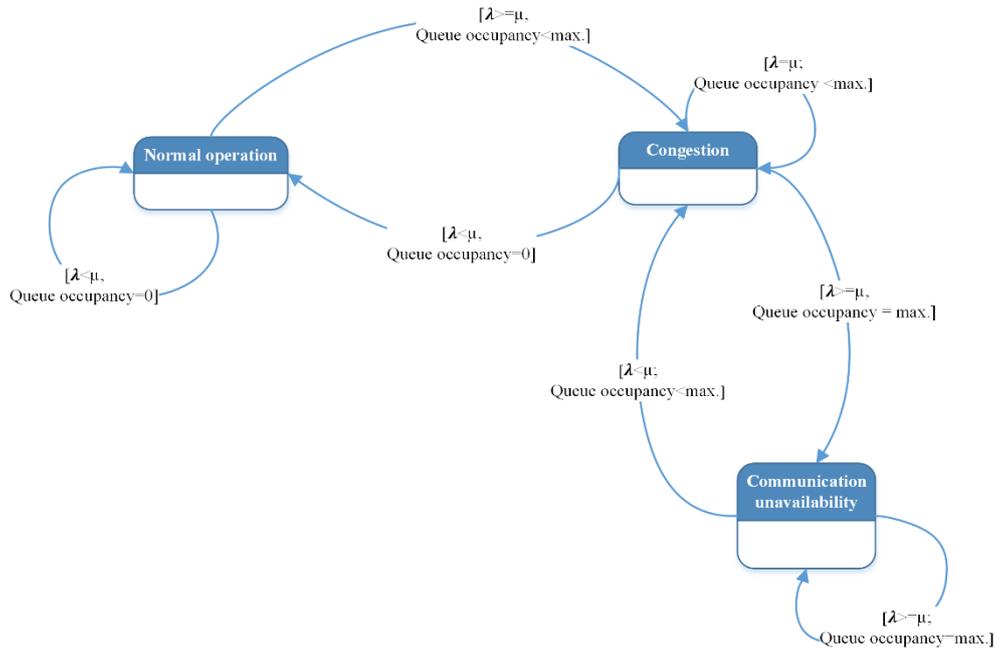

Figure 2 - UML diagram of the state of IC resources and services [34]

A DDoS attack exploits the shortcomings of the Internet network architecture. An Internet network is designed to provide functionality where security of communication is not considered. Studies [8, 35, 36] highlight the shortcomings of the Internet network design that enable successful implementation of DDoS attacks. The first disadvantage is reflected in the limited capacity of networked devices that can be completely depleted. Also, as one of the problems, it is pointed out that the vulnerability to a DDoS attack does not depend on the level of protection of the devices in the network that is the target of the attack. It depends on the security of the global Internet network. The increase in the number of devices and the amount of traffic transmitted through the network results in the upgrade of high-bandwidth links in the communication network core segment. This makes it possible to use such a configuration to deliver a traffic intensity exceeding 1Tbps towards the attack ultimate target. Failure to authenticate an individual IP packet enables the implementation of DDoS attacks with fake source IP addresses, which makes a strong mechanism for anonymizing the source of the attack. Finally, one of the more significant problems is the distributed management and control methods in the Internet network. Any local network connected to the Internet can operate according to the policy defined by the administrator. Consequently, there is no way to implement and enforce global security mechanisms or policies that make it impossible to investigate traffic behavior between diverse networks.

### 3.2   Impact of botnets on the generation of DDoS attacks

The concept of a botnet as a network of remotely managed vulnerable devices has been known for more than 20 years. However, its application in the IoT concept environment is relatively new. Creating a botnet requires the implementation of malicious software in a target device that will allow remote management. According to [37–40] due to the known limitations of IoT devices, many malicious software are aimed at this type of device in order to create a botnet network through which it is possible to conduct DDoS attacks that generate DDoS traffic of high intensity (above 1Tbps) [41]. According to research by Bitdefender, in 2018, 78% of illegitimate activity was led by IoT botnets [42]. Although the possibility of forming a botnet using IoT devices has been known since 2009, when the first such DDoS attack was performed, it was in 2016 that there was the first publicly exposed DDoS botnet attack called Mirai [43]. Mirai is malicious software from the class of computer worms that detects vulnerable devices in the network, exploits their vulnerabilities to install malicious code and further propagates through the network. Several studies have analyzed the Mirai malware and botnet network in order to discover how it works and the effects on devices and the DDoS traffic generated through that botnet. Research [44] shows the working principle of the Mirai botnet, shown in Figure 3.

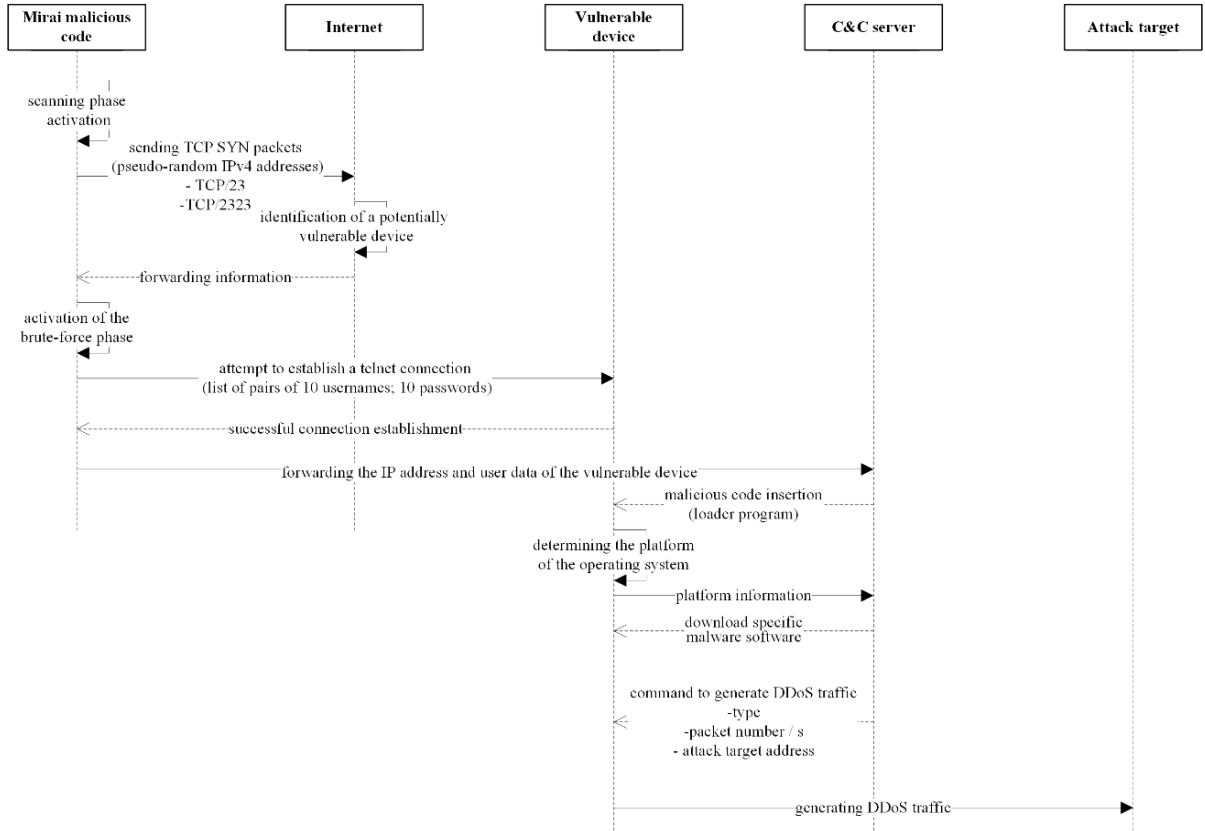

Figure 3 - Sequential UML diagram of the Mirai botnet working principle [44]

According to Figure 3, Mirai malware works in two phases: scanning and brute-force. In the implementation of the first phase, the pseudorandom range of IPv4 addresses is scanned using the TCP SYN method on the communication ports used by the telnet network service (TCP 23 and TCP 2323). After an active IoT device is detected in the network that has an open communication port 23 or 2323, the second phase of operation begins.

The second phase uses a brute-force attempt to illegally access a vulnerable device via a telnet service and a list of 62 potential usernames and passwords. After successfully accessing the device, Mirai sends the device IP address and the associated access data to the control server. The control server implements a program code (loader program) in the vulnerable device, whose role is to detect the device operating system based on which the malicious software specific to the identified operating system is implemented. Implemented malicious code allows the control server to remotely control the device in the form of forwarding instructions for generating DDoS traffic (type of attack, traffic intensity, protocol, target of the attack) [44,45].

According to research [37], the Mirai botnet consisted of approximately 500,000 IoT devices from 164 countries whose purpose was to generate DDoS traffic to many destinations, the most prominent of which were the KerbsOnSecurity.com web server, the Dyn DNS service and the French web hosting provider OVH [43, 46]. The characteristics of DDoS attacks using the Mirai botnet are shown in Table 1.

Table 1 - The most significant DDoS attacks conducted through the Mirai botnet

| Attack target | Number of devices | DDoS traffic intensity | Botnet |
| --- | --- | --- | --- |
| KrebsOnSecurity.com | 24,000 | 623 Gb/s | Mirai |
| Dyn DNS | 100,000 | 1,2 Tb/s | Mirai |
| OVH | 145,607 | 1,1 – 1,5 Tb/s | Mirai/Bashlight |

In addition to the Mirai botnet, as the most important representative and indicator of the importance and impact of DDoS attacks based on SHIoT devices, research also indicates many other botnets that use SHIoT devices to generate DDoS traffic [37, 43, 45–49].

Table 2 shows the existing malicious software whose purpose is to create a SHIoT botnet to generate DDoS traffic. In addition to malicious software, the types of DDoS traffic are listed in terms of the protocols used that individual botnets can generate.

Table 2 - Malicious software used to create an IoT botnet to generate DDoS traffic [46]

| Malicious software | Type of generated DDoS traffic |
| --- | --- |
| Linux.Hydra | SYN Flood, UDP Flood |
| Psyb0t | SYN Flood, UDP Flood, ICMP Flood |
| Tsunami, Kaiten | SYN Flood, UDP Flood, ACK Flood |
| Aidra, LightAidra, Zendran | SYN Flood, UDP Flood, ACK-PUSH Flood, HTTP Layer 7 Flood, TCP XMAS |
| Spike, Dofloo, MrBlack, Wrkatk, Sotdas, AES.DDoS | SYN Flood, ACK Flood |
| BASHLITE, Lizkebab, | SYN Flood, UDP Flood, ICMP Flood, DNS Query Flood, HTTP Layer 7 Flood |
| Elknot, BillGates | SYN Flood, UDP Flood, ACK Flood |
| XOR.DDoS | SYN Flood, UDP Flood, ICMP Flood, Flood, DNS Amplification, HTTP Layer 7 Flood, other TCP Floods |
| LUABOT | HTTP Layer 7 Flood |
| Remaiten, KTN-RM | SYN Flood, UDP Flood, ACK Flood, HTTP Layer 7 Flood |
| NewAidra, Linux.IRCTelne | SYN Flood, ACK Flood, ACK-PUSH Flood, TCP XMAS, TCP Floods |
| Mirai | SYN Flood, UDP Flood, ACK Flood, VSE Query Flood, DNS Water Torture, GRE IP Flood, GRE ETH Flood, HTTP Layer 7 Flood |

From all the above, it is possible to conclude that, generically speaking, DDoS attacks and thus DDoS traffic as a product of such attacks are a growing security threat to IC resources and service availability. This is confirmed by numerous studies that indicate an increase in the sophistication of this threat in terms of the diversity of protocols used to generate DDoS traffic.

An additional factor in the growth of this problem is undoubtedly the emergence and growth and development of the concept of smart home and IoT devices which due to their limitations and security shortcomings allow the creation of botnets with a much larger number of remotely controlled devices that have the potential to generate DDoS traffic.

### 3.3   Analysis of COVID-19 pandemic impact on the e-learning system

Despite continuous research into network traffic anomalies, cyber attacks such as DDoS attacks are still frequent and can have numerous negative effects on the IC system predicted performance and the availability of IC services. The pandemic of coronavirus (SARS-CoV-2) highlighted the importance of the availability of e-learning systems and services. Crises such as the mentioned pandemic result in the need for isolation of users (students/teachers/administrators) whereby education and the supporting processes rely on the reliable work of the e-learning system and all its elements. On March 16th, 2020, a DDoS attack on AAI@EduHr system was responsible for authenticating the users to access various e-learning services (Merlin, webinar, e-learning center, *Dabar*, *Hrčak*, filesender and others) in the Republic of Croatia [3]. The conducted attack indicated the need to research and find solutions for cyber threats (primarily DDoS) applicable in specific scenarios.

According to data from the Croatian University Computing Center (*Srce*) during March 2020, in the month in which remote education began due to the COVID-19 pandemic, the AAI@EduHr system recorded 11,214,236 successful authentications for 517,453 unique users (Figure 4). For comparison, in March 2019, the AAI@EduHr system had a total of 3,220,212 successful authentications of 252,974 unique users, which can be seen in Figure 5. Authentication through AAI@EduHr was most commonly used to access MS Office 365 systems for schools, Loomen, and *Srce* systems: Merlin - a remote learning system for students and teachers ISVU Information System for Higher Education (primary the *Studomat* module) [50]. In two and a half days trough e-learning system Merlin *webinar* 420 lectures longer than 15 minutes were held with 2,594 participants.

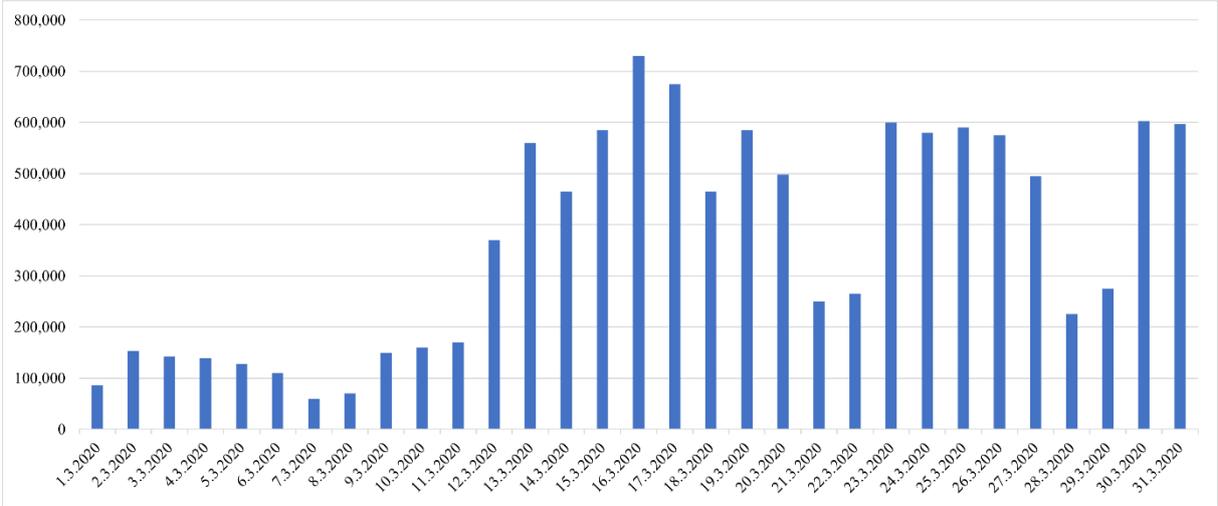

Figure 4 - Number of successful authentications in March 2020 by day [50]

The presented data indicate the occurrence of the flash crowd phenomenon. This occurs when legitimate requests to access a web-based service exceed the statistically normal number of legitimate requests [51]. Accordingly, flash crowds may adversely affect the performance of DDoS attack detection models based on machine learning methods that use datasets created during the period when the number of service requests is common. Such models are often detecting flash crowd traffic as a DDoS attack traffic, even though it represents a legitimate service request. Some authors research solutions that can differentiate flash crowd and DDoS traffic by using human behavior and interaction, but such solutions are not acceptable to users [52].

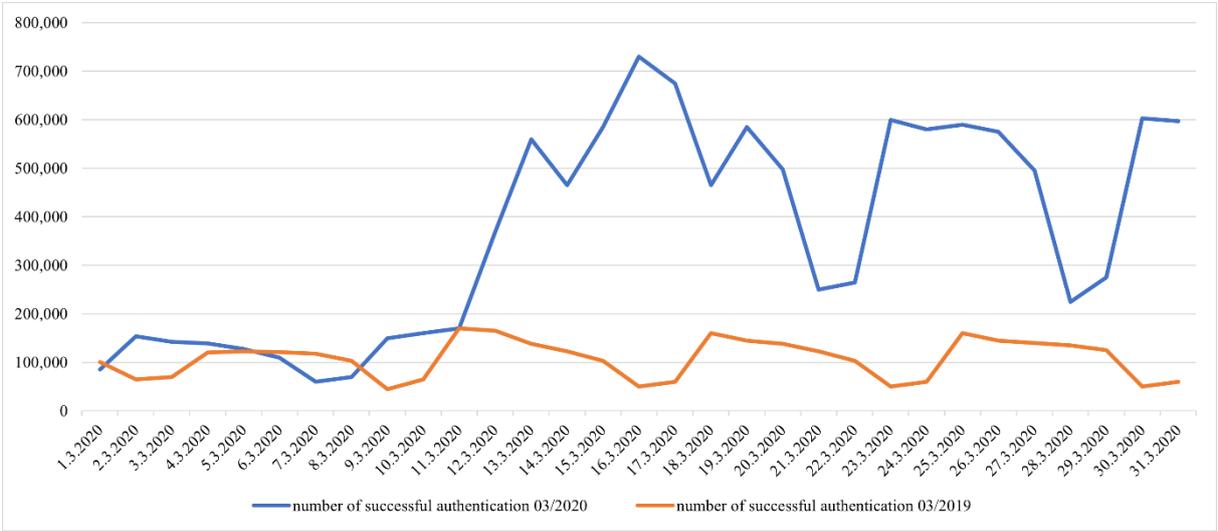

Figure 5 - Number of successful authentications of March 2019 and 2020 comparison [50]

To achieve effective detection of DDoS and thus for reaction to such attacks in specific cases of flash crowd phenomena under crises such as a coronavirus pandemic, the detection model should be based on distinguishing between DDoS traffic and traffic generated under flash crowd conditions. One of the research directions followed by this project proposal is identifying the unique characteristic of flash crowd traffic vis-à-vis DDoS traffic as a basis for the development of DDoS traffic detection model [53].

# 4 Methodology for developing DDoS traffic detection model

The implementation of efficient and effective research for dealing with the described problem requires well-structured research methodology [54].

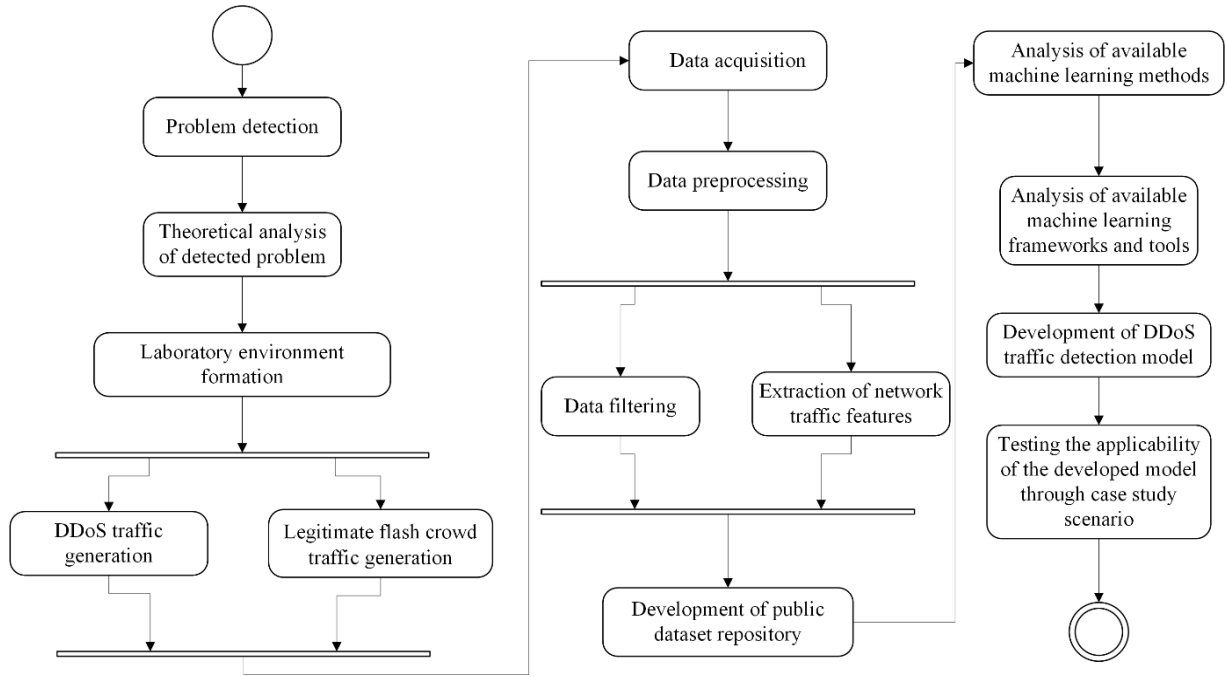

Figure 6 - UML activity diagram of activities through research phases [54]

Research implementation through four-phase methodology, and the activities shown in Figure 6 are proposed as follows: (1) analysis of the existing elements of the e-learning ecosystem and establishing the theoretical basis and differences between DDoS and flash crowd traffic, (2) forming of a laboratory environment, the collection of network traffic, processing and preparation of the collected data for further analysis, (3) development of DDoS traffic detection model, validation and performance evaluation of the developed DDoS traffic detection model, (4) analysis of the applicability of the developed anomaly detection model and the reaction capabilities based on the operation of the developed model.

## 4.1 Establishing the theoretical basis and differences between DDoS and flash crowd traffic

In the first phase of the research the current scientific literature will be analyzed to identify the elements of the e-learning ecosystem, architecture, communication topologies and technologies, and other relevant characteristics of such environments. The purpose of the aforementioned research activities is to provide adequate recommendations and guidelines in the last phase of the research to ensure the availability of e-learning services and minimize the negative impacts of the emergence and realization of cyber threats such as DDoS attacks. The current scientific literature analysis will also determine the current research findings related to the characteristics of the network traffic generated as a result of flash crowd activity and current research achievements regarding the development of models and systems that can distinguish DDoS and flash crowd traffic.

## 4.2 Formation of laboratory environment and data collection

Second phase should include establishing a laboratory environment for generating and collecting legitimate (flash crowd) and illegitimate (DDoS) traffic. The laboratory environment is planned to be established at the Department of Information and Communication Traffic at the FPZ within the Laboratory for Security and Forensic Analysis of the Information and Communication System (LSF). Flash crowd and DDoS traffic generation is planned with the implementation of the dedicated hardware platform. Such a platform can simulate realistic flash crowd and DDoS traffic and provide advanced management and define features of the generated traffic at different OSI layers (Open Systems Interconnection) model. The next planned activity is to collect and store the generated traffic on the dedicated computers in .pcap format files, which is suitable for further manipulation in the form of analysis and extraction of the traffic feature values.

Further research activities in this phase are related to the pre-processing and preparation of data for further analysis. This implies filtering the collected data and extraction of the network traffic features that can be used in network anomaly detection model and adequate labeling of feature vectors. Considered will be exclusively the packet header values, i.e. statistical features of traffic while the packet content will not be considered because of the application of cryptographic methods in the communication processes. In this phase, the independent traffic features will be selected based on which it is possible to differentiate the observed traffic on legitimate and DDoS. Also, the level of connection with the dependent feature will be determined and those independent features with the highest level of connection with the dependent feature will be selected. The selected features will be used to define the model in the next research phase. The final output of the second phase will be the datasets containing legitimate traffic generated as a result of flash crowd activities on the e-learning ecosystem elements (simulated through dedicated network generator hardware) and DDoS traffic as a network traffic anomaly generated as a result of DDoS attack (simulated through dedicated network traffic generator hardware). By collecting the generated legitimate and illegitimate traffic, a unique dataset will be formed. Such a dataset will be published on a publicly available server (on FPZ servers) in original and in the processed form for the academic community to further research detection and reaction on DDoS attacks in specific scenarios.

### 4.3 Development of DDoS traffic detection model

The previously carried out activities laid the foundation for the third phase of the planned research. This phase encompasses the development of the network traffic anomalies detection model. The first step in this phase is to analyze and select the adequate supervised machine learning method from the set of ensemble methods that will suit the solving of binary classification problem. In solving such a problem, the objective is to check the generated new traffic sample congruence with the sample of the legitimate traffic. The incongruity of the traffic feature values with the values of legitimate traffic features above the defined threshold will mean that the device generates DDoS traffic within the observed time. The following activity is the development of network anomaly detection model using the supervised machine learning methods. For the implementation of the chosen method available machine learning platform will be used such as TensorFlow, Weka, KNIME, Orange, Hadoop, Apache Spark, Neo4ji, R or similar. The last step in this phase is to validate the model performances through standard validation measures for the classification models (accuracy, precision, specificity, kappa coefficient, rate of false and true positive results, confusion matrix, etc.).

### 4.4 Applicability of the developed anomaly detection model

In the final, fourth research phase, the developed model applicability in real scenario through case study will be analyzed. The guidelines and recommendations for the response to the anomalies detected with the developed model will be defined. The purpose of this activity is to minimize the negative effects of cyber threats such as DDoS on the availability of e-learning services particularly in the crises such as pandemic of SARS-CoV-2 virus when such services are critical for the undisturbed running of the educational and supporting processes.

## 5 Discussion and conclusion

Under the coronavirus pandemic influence, the e-learning systems have become a model for implementing the teaching process with no alternative. In such crisis circumstances, e-learning system availability is emphasized, which can be very easily reduced by using DDoS attacks. DDoS attacks are easy to implement, and their effectiveness is high. As the number of users requiring a particular resource increases, so does the risk that malicious users will carry out DDoS attacks that will reduce or completely disable such system availability to legitimate users. Kaspersky's report also supports this thesis, pointing out that in Q1 2020, the number of DDoS attacks on the e-learning systems increased by 350% compared to the same period in 2019. Also, the importance of the e-learning system availability can be seen from the example of a DDoS attack on the authentication system of the e-learning system users (AAI@EduHr) in the Republic of Croatia at the time when the entire educational process was transferred online. The specificity of the observed scenario of the implementation of DDoS attacks on the e-learning systems in the pandemic circumstances is reflected in the sudden increase in the demand by a large number of users for e-learning services. That is a phenomenon called the flash crowd event, representing a significant increase in the number of legitimate requests. When a flash crowd event and a DDoS attack coincide, a specific scenario is created that requires models and systems that can distinguish between two types of generated requests and network traffic that are approximately the same in terms of characteristics.

In this paper, we proposed a research methodology customized for finding solutions to preserve the availability of e-learning systems from DDoS attacks in the crises in which such systems have no alternative, such as the current coronavirus pandemic. Using the proposed methodology for developing DDoS detection model and guidelines and recommendations for the reaction to the detected network anomalies provide the potential for further practical application and substantial socio-economic benefits from several aspects. The research implementation within the framework of the proposed project is significant for developing the research area since it considers the challenges in a specific scenario of using e-learning services resulting from the COVID-19 pandemic. The proposed research is planned to form open and public dataset repositories containing network traffic (in raw and preprocessed format) generated by simulating flash crowd scenarios and DDoS attacks. Such repository will benefit other researchers for further research of behaving flash crowd traffic and distinguish it from other types of traffic such as DDoS or some other types of traffic anomalies. The planned extensive use of machine learning, especially ensemble type of machine learning methods, will result in a developed anomaly detection model that can proactively detect cyber threat and adequately react to such a threat. In that way, it will give the possibility to secure the availability of e-learning ecosystems and services when needed most, and that is in crisis scenarios such as pandemic of coronavirus and similar. Securing the availability of e-learning services that are becoming critical will increase user satisfaction on every level (students/teachers/administrators) and allow undisturbed running of educational and supporting processes. Furthermore, this kind of research would open a variety of related research problems and it would enable knowledge transfer between diverse research teams and e-learning system operators as a response to high-risk situations and crisis scenarios that may occur in the future.

Future research is planned to implement the proposed methodology and develop a cyber intrusion detection model at the conceptual, prototype, and pilot levels. Also, future research is planned to identify the possibilities of protecting e-learning systems from the detected cyber threats.